\documentclass[fleqn,10pt]{wlpeerj}

\usepackage{hyperref}

\title{Human-aligned quantification of numerical data}

\author[1,2]{Anton Kolonin}
\affil[1]{The Artificial Intelligence Research Center Novosibirsk State University, Novosibirsk, 630090, Russia}
\affil[2]{SingularityNET Foundation, Baarerstrasse 141, 6300 Zug, Switzerland}
\corrauthor[1]{Anton Kolonin}{akolonin@gmail.com}

\begin{abstract}
Quantifying numerical data involves addressing two key challenges: first, determining whether the data can be naturally quantified, and second, identifying the numerical intervals or ranges of values that correspond to specific value classes, referred to as "quantums," which represent statistically meaningful states. If such quantification is feasible, continuous streams of numerical data can be transformed into sequences of "symbols" that reflect the states of the system described by the measured parameter. People often perform this task intuitively, relying on common sense or practical experience, while information theory and computer science offer computable metrics for this purpose. In this study, we assess the applicability of metrics based on information compression and the Silhouette coefficient for quantifying numerical data. We also investigate the extent to which these metrics correlate with one another and with what is commonly referred to as "human intuition." Our findings suggest that the ability to classify numeric data values into distinct categories is associated with a Silhouette coefficient above 0.65 and a Dip Test below 0.5; otherwise, the data can be treated as following a unimodal normal distribution. Furthermore, when quantification is possible, the Silhouette coefficient appears to align more closely with human intuition than the "normalized centroid distance" method derived from information compression perspective.
\end{abstract}

\begin{document}

\flushbottom
\maketitle
\thispagestyle{empty}

\section*{Introduction}

When developing analytical and predictive applications that rely on numerical data, it is often useful to convert streams of numbers into sequences of symbols. This transformation simplifies the input data, making it more efficient to process and allowing for tokenization into chains of specific states represented by the data. For instance, this approach can facilitate the logical analysis based on the data according to \cite{Boros1997} by applying symbolic logic to continuous streams of numerical data that describe specific properties of objects, provided these properties can be identified by a finite set of characteristic values with certain distributions around them. This process may involve identifying "natural" ranges of numerical values or "classes" (also referred to as "quantum" states or "quantums") associated with particular states of the system or process being measured by a specific data source.

This process can be referred to as "quantification" or "binarization" according to  \cite{Clark1976} and is particularly applicable when the distribution of data values is multimodal rather than unimodal, see \cite{6137392}. For instance, if the data can be characterized as a bimodal distribution, it can be represented as a binary stream where 0 or 1 corresponds to one of the two modes, each representing a specific state (such as the average selling price or buying price in a stock exchange). For another example, multimodal distributions—like varying traffic speeds that correspond to different traffic regulations and vehicle equipment—can result in multiple modes that represent different system states (for example, "pedestrian zone speed," "city speed," and "highway speed"). In each of these instances, when a stream of numerical data can be represented as a sequence of distinct states, each state can be designated as a "symbol." This enables the use of symbolic or predicate logic methods to analyze the data.

Potential practical applications of the aforementioned quantification include the implementation of causal analytics for financial predictions and decision support systems as presented in \cite{10.5555/342932} and \cite{KOLONIN2022180}, as well as natural classification systems for different domains, see \cite{VITYAEV2015169}.

In this study, we approach the data quantification problem as a clustering problem in a one-dimensional space with a dimension corresponding to a specific function or measurable property supported by the data. In future research, we plan to extend the approach to multidimensional data, such as multi-parameter market dynamics forecasting according to \cite{10.5555/342932} and \cite{10.1007/978-3-031-19907-3_4} as well as unsupervised natural language learning following \cite{10.1007/978-3-030-27005-6_11}.

Since the goal of process automation is to reduce human labor or to assist human personnel, we believe that metrics should match human values, work experience, and mental models. We study how different metrics fit with human perspectives based on alignment metrics such as Fleiss' kappa and Krippendorff's alpha, according to \cite{Zapf2016}.

For clustering purposes, we considered using the widely used K-means method, clustering parameterized by the target number of clusters $K$ according to  \cite{Jin2010}, and the DBSCAN algorithm parameterized by the $epsilon$ value following \cite{9356727}. Any clustering algorithm has such a parameter to impact on the expected number of clusters explicitly, as in the case of K-means, or implicitly, as in the case of DBSCAN, so the role of the human is to tune the parameter according to the data and business goals.

The goal of our work is to explore to what extent we can delegate this particular human role to an algorithm and what tools and metrics we can use to do this.

One of the algorithms is the Dip Test, which is used to check whether a distribution can be described as “normal” or not, according to earlier studies according to \cite{10.1214/aos/1176346577} and \cite{Bauer_2023}. If it is not a “normal” (unimodal) distribution, then we can try to split it into several distributions corresponding to the “modes” of the multimodal distribution.

Another tool is the Silhouette coefficient, according to \cite{4426662}, which can be used to find the optimal placement of data points into clusters with minimal overlap and conflicts. The problem with this tool is that by definition it is only applicable for a number of clusters greater than or equal to two, so it can help choose between bimodal and trimodal distributions and more multimodal ones, but it cannot choose between unimodal and multimodal distributions in general.

\begin{figure}[ht!]
  \centering
  \includegraphics[width=0.9\textwidth]{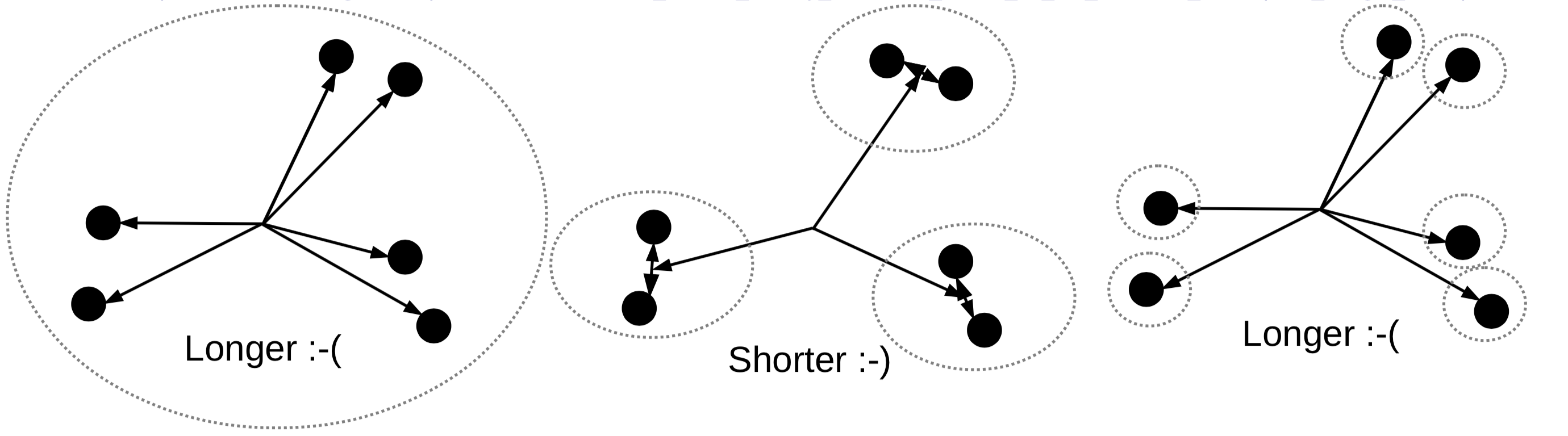}
  \caption{Example of using the "Normalized Centroid Distance" metric: Left - one cluster, 6 distances from one centroid, the total length is large; Center - three clusters, 6 distances for each of the 3 centroids to 2 data points in each plus 3 distances from the centroid to the center, the total length is the smallest; Right - six clusters, 6 distances from the common center of the centroids plus 6 zero distances from each centroid to every 1 point in its cluster, the total length is large. The optimal number of clusters based on the minimum total distance is 3.}
  \label{fig1}
\end{figure}

The alternative approach we explore is inspired by the "Minimum Description Length Principle" according to \cite{10.1007/978-3-642-23851-2_9} and recent work finding that practically sensible data structures used for communications, such as human languages, are based on the efficiency of information compression criteria, as shown by \cite{10.1007/978-3-031-44865-2_1} and \cite{pagnoni2024bytelatenttransformerpatches}. Building on these works, we explore the use of a metric called "Normalized Centroid Distance" that measures the total length of the paths needed to connect all input "real" data points into a single graph, including all "synthetic" points corresponding to the centroids of all clusters describing the data. An example is shown in the Figure~\ref{fig1} for the case of two-dimensional clustering for different numbers of clusters describing the same data points. In each of the three clustering options, we calculate the total distances from the cluster centroids to all data points in the corresponding clusters, as well as the distances between the centroids of each cluster to the mean center of all centroids. We expect that the optimal clustering option will be determined by the minimum total distance.

\section*{Experimental Study}

\subsection*{Methods}

To evaluate the methods described above, we used one-dimensional synthetic data generated from exponential-logarithmic, harmonic, and normal distributions with the number of distribution modes from 1 to 3. The most representative results presented in this section were obtained using normal distributions with different numbers of modes. To calculate the normal distributions of multimodal data, we used the $make{\_}blob$ function described by \cite{Sierra-Sosa2023}. For the following experiments presented, we used three one-dimensional "blob" centers with values 1, 4, and 5 corresponding to the respective modes. That is, generating synthetic data with a standard deviation "std" of 0 would give us data with only three discrete "quantum" states at 1, 4, and 5, respectively.

\begin{figure}[ht]
  \centering
  \includegraphics[width=0.9\textwidth]{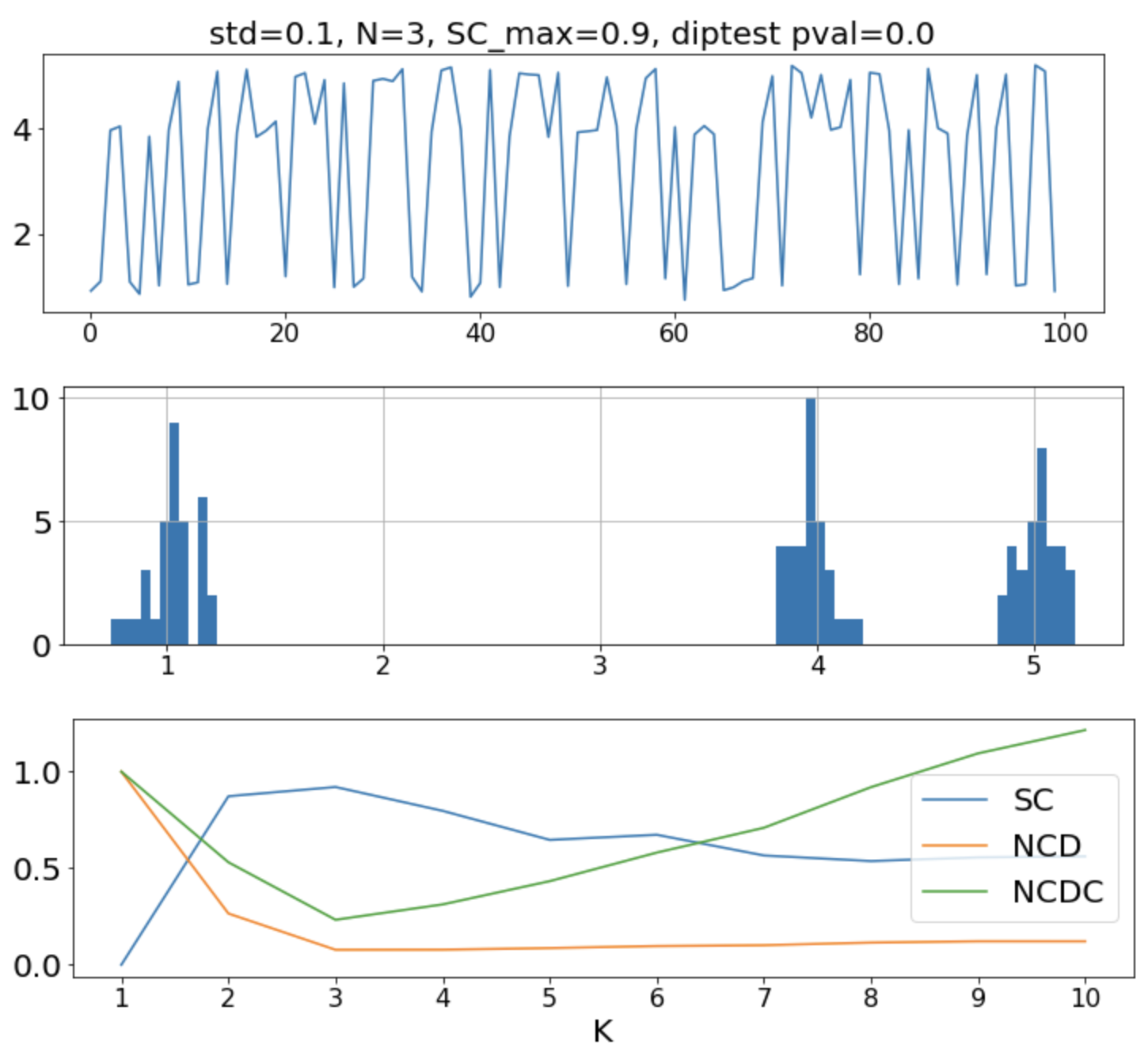}
  \caption{The original input data are 100 points for $std = 0.1$ (top), corresponding to the obvious 3 distribution modes (middle), and plots for metrics calculated for different numbers of clusters with $SC$ and $NCDC$ agreement at $K = 3$ (bottom).}
  \label{fig2}
\end{figure}

\begin{figure}[ht]
  \centering
  \includegraphics[width=0.9\textwidth]{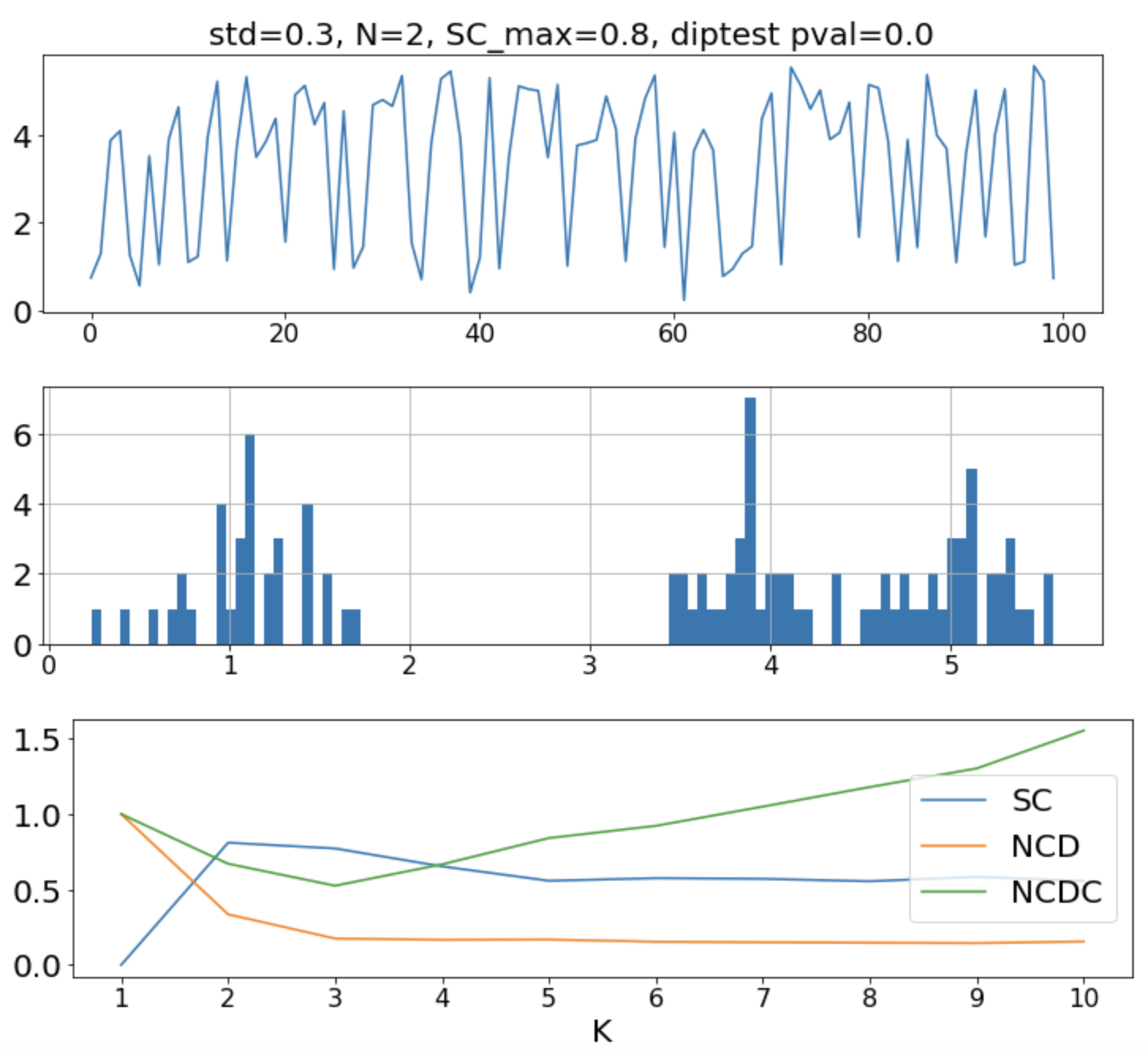}
  \caption{The original input data are 100 points for $std = 0.3$ (top), corresponding to the obvious 2 and possibly 3 distribution modes (middle), and plots for metrics calculated for different numbers of clusters with maximum $SC$ for $K = 2$ and minimum $NCDC$ for $K = 3$ (bottom).}
  \label{fig3}
\end{figure}

\begin{figure}[ht]
  \centering
  \includegraphics[width=0.9\textwidth]{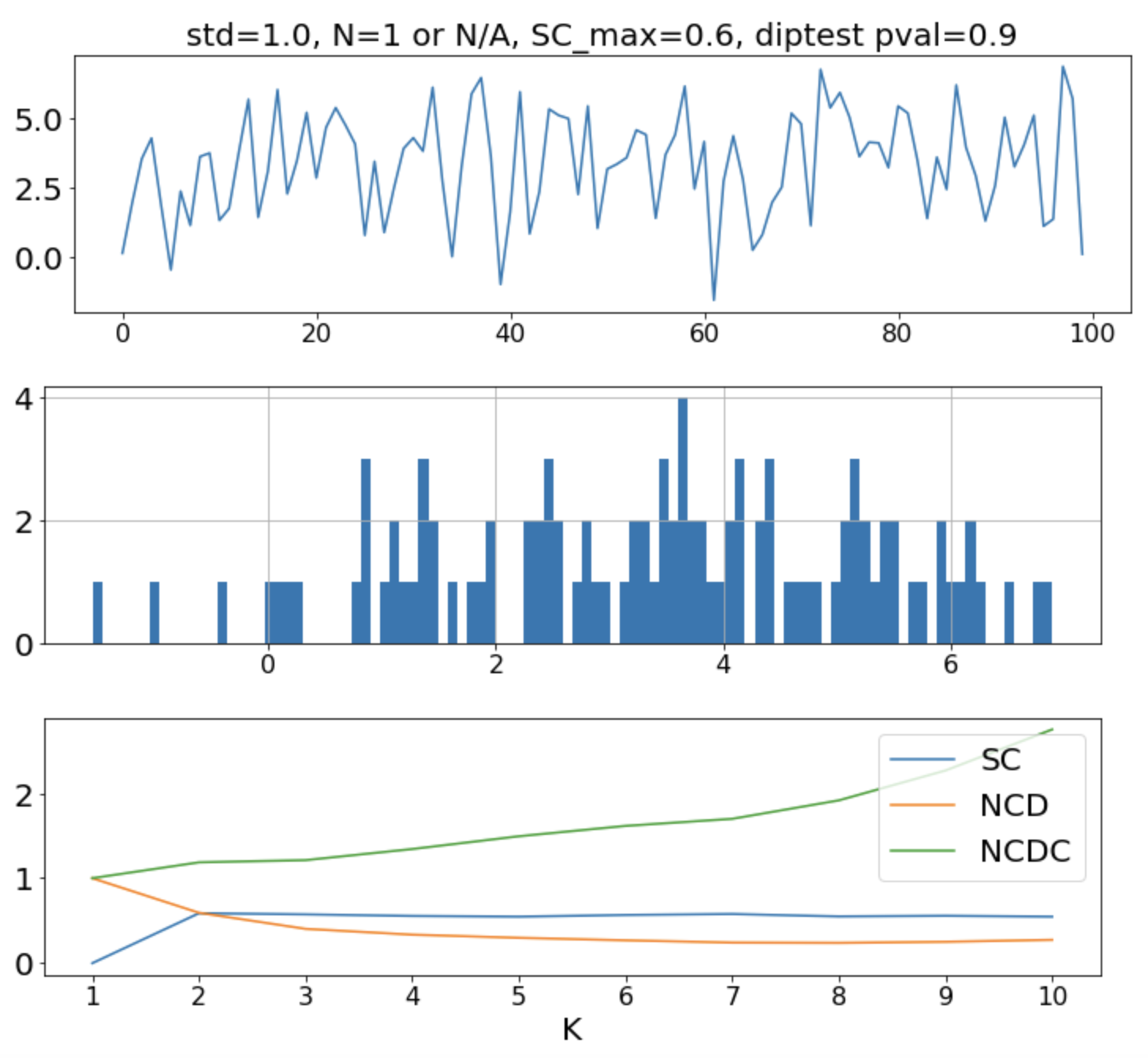}
  \caption{The original input data are 100 points for $std = 1.0$ (top), corresponding to a single apparent mode or normal distribution (middle), and plots for metrics calculated for different numbers of clusters without an expressive $SC$ maximum, with $NCDC$ minimum for $K = 1$ (bottom).}
  \label{fig4}
\end{figure}

In order to generate different "realistic" non-discrete distributions with different numbers of modes, we used non-zero values of $std$, so that $std = 0.1$ corresponded to the expected 3 clusters with corresponding one-dimensional centroid coordinates at 1, 4, and 5 (Figure~\ref{fig2}). Furthermore, setting $std = 0.3$ created an apparent separation of data points in two clusters with corresponding centroids at 1 and 4.5 due to the large overlap of individual distributions at 4 and 5 (Figure~\ref{fig3}). Finally, having $std = 1.0$, we obtained a complete overlap of all three original distributions, which would be better described by a single effective distribution with only one centroid (Figure~\ref{fig4}). All experiments were performed with different numbers of data points - 100 and 1000 and different values of $std$. The following three figures show the most typical results.

After initial experiments using the K-means and DBSCAN clustering algorithms, we decided to use K-means because the parameter $K$ that determines its performance is explicitly related to the number of clusters, which is not the case with DBSCAN, where the parameter $epsilion$ affects the number of clusters implicitly.

For each data set with a corresponding $std$ value, we performed clustering with a different parameter $K$, computing the following metrics for each set of $K$ clusters with respect to the current set of data points.  

Since in all our test settings the number of expected distribution modes varied from 1 to 3, we decided to limit the range of $K$ under study to 10, which was confirmed by the global extremes of all the metrics under study in this range, as we will see further.

To confirm the reproducibility and robustness, given the knowledge of the stochastic nature of the K-means algorithm, we performed multiple runs with different numbers of data points (100 and 1000) and different values of the K-means $random_state$ parameters, such as None, 1, 2, and 3.

\subsection*{Computable Metrics}

$SC$ - "Silhouette coefficient", according to \cite{4426662} was calculated for each $K$ and was set to 0 in case $K=1$ (since it is non-computable at this point by definition). The optimal value of $SC$ by definition corresponds to its global maximum.

$NCD$ - "Normalized Centroid Distance" as the normalized centroid distance computed from the data of $N$ points $V_i$ as follows. First, we compute the mean center of the distribution $M$. Then we compute the baseline distance $D_1$ for one cluster $K=1$ as the sum of the distances of all data points $V_i$ from $M$. Then, for each given number of clusters $k$, we compute $D_k$ as the sum of the distances between the data points $V_i$ and the corresponding clusters $C_i$ plus the sum of the distances of each cluster centroid $C_j$ from the mean $M$. Finally, we normalize $D_k$ by $D_1$. The optimal value of $NCD_k$ will by definition correspond to its global minimum, as shown in the example of the three clusters in the center of Figure~\ref{fig1} above. However, we found that using this metric in the case of our multimodal but one-dimensional data data is difficult, since it usually has a very broad and weakly defined global minimum in the form of a "long tail" in the region with $K$ greater than the reference $K = 3$. Thus, no meaningful global minimum can be determined or it can be mistakenly determined with a larger number of clusters or modes than expected.

\[
M = \frac{\sum_{i}V_i}{N}
\]
\[
D_1 = \sum_{i}V_i-M
\]
\[
D_k = \sum_{i=1,N}V_i-C_i + \sum_{j=1,k}C_j-M
\]
\[
NCD_k = D_k / D_1
\]

$NCDC$ - "Normalized Centroid Distance times Centroids" as the normalized centroid distance multiplied by the number of clusters. This metric was proposed due to the weakness of the $NCD$ metric discussed above. We devised the $NCDC$ metric with an additional factor penalizing the increase in the number of clusters or modes, so that we could find a meaningful global minimum corresponding to the optimal number of clusters.

\[
NCDC_k = k * D_k / D_1
\]

$SC+$ is a combined metric based on the rule that $SC$ is used to determine $K$ if its maximum value exceeds a threshold of 0.65; otherwise, the $NCDC$ metric is used. This was proposed due to the observation that the basic $SC$ is not computable at $K = 1$, while distributions close to normal tend to have $SC$ below 0.65. At the same time, $NCDC$ is computable at any $K$, including $K = 1$, when DipTest $pval$ is close to 1.0, while the distribution is close to unimodal or normal in particular.

Additionally, for each dataset, we computed the DipTest metric $pval$ according to earlier studies by \cite{10.1214/aos/1176346577} and \cite{Bauer_2023}) to assess the "normality" of the distribution, assuming that if it is not "normal", then it can be considered multimodal.

\subsection*{Results and Discussion}

The results are illustrated for std values of 0.1 (obviously 3 clusters), 0.3 (probably 2 clusters or possibly 3 clusters), and 1.0 (no clusters or 1 cluster, which is the same) in Figure~\ref{fig2}, Figure~\ref{fig3}, and Figure~\ref{fig4}, respectively.

In the case of $std = 0.1$, when 3 clusters are clearly distinguishable (Figure~\ref{fig2}), the Dip Test $pval$ is expectedly equal to zero, which indicates that the normal distribution cannot be applied, $SC$ reaches its expected maximum of 0.9 at $K = 3$, $NCD$ decreases sharply to a point at the expected $K = 3$, but then continues to decrease slowly to the point $K = 9$, which is not expected. In turn, $NCDC$ has a clear global minimum at $K = 3$. That is, at least the three metrics $pval$, $SC$ and $NCDC$ agree on the presence of a non-normal trimodal distribution.

In the case of $std = 0.3$, where there is room for an argument if there are 2 or 3 clusters (Figure~\ref{fig3}), the Dip Test $pval$ is expectedly equal to zero, but $SC$, $NCD$, and $NCDC$ suggest different numbers of $K$. $SC$ is confident at $K = 2$ with its global maximum value of 0.8, while $NCDC$ reaches its global minimum at $K = 3$. Again, as in the previous case, $NCD$ reaches its minimum at the distant point of $K = 9$. In the next section, we will discuss to what extent this kind of disagreement is aligned with human view.

In the case of $std = 1.0$, when it is impossible to determine a reliable number of clusters or one could say that there is only one cluster (Figure~\ref{fig4}), the Dip Test $pval$ is expectedly close to 1.0 (the actual value is 0.9), indicating a probable normal distribution. Moreover, $SC$ does not provide any discernible maximum, having a value nearby 0.6 for all $K$ numbers. In its turn, as in the previous cases, $NCD$ gradually decays to greater number of $K$, suggesting the creation of $K = 8$ clusters. Finally, $NCDC$ correctly reaches a global minimum at $K = 1$, in accordance with the Dip Test estimate, indicating only one distribution mode.

\section*{Human Evaluation}

\subsection*{Methods}

To assess the extent to which the various metrics described above can help in finding clustering parameters that match human common sense and intuition, we conducted a field study. The study was conducted as a one-time, non-cohort, anonymous study. Participants were members of a professional online data science community who were willing to anonymously complete an online Google Form. The study involved anonymized data labeling as distributions generated from synthetic one-dimensional numerical data and presented as distribution histograms according to a proposed questionnaire. The data from the form were collected over a period of 1 month. The study was approved by Institutional Review Board "The Ethics Committee Of Novosibirsk State University", in Protocol No. 5, dated December 23, 2024. In particular, the need for written consent from participants was waived by the board based on the following statement: "The survey is conducted through an anonymous questionnaire via the Google Forms online tool without collecting any identifying information about the participant. The form and composition of the questionnaire does not imply the receipt of any personal information, but only contains a subjective visual assessment of the number of distribution modes (one-dimensional clusters) by the anonymous participant. In connection with the above, obtaining de-anonymized personal informed consent is not required." Because of this, no informed consent was collected in this anonymized study.

During the study, 14 respondents filled the form anonymously. Each respondent was provided with a series of distributions of different types - unimodal or multimodal, exponential logarithmic or harmonic - for different numbers of data points (100 and 1000). Each item in the series contained four different data distributions of the same type for the same number of data points. For each distribution, the respondent had to choose how many clusters or distribution modes he or she sees: 1 (no multimodality at all), 2, 3, or 4 or more. Independently of human judgments, we chose the number of clusters $K$ based on the $NCDC$ and $SC+$ metrics defined earlier. The most representative items for the questionnaire are presented in Figure~\ref{fig5}.

\subsection*{Results and Discussion}

\begin{figure}[ht!]
  \centering
  \includegraphics[width=0.9\textwidth]{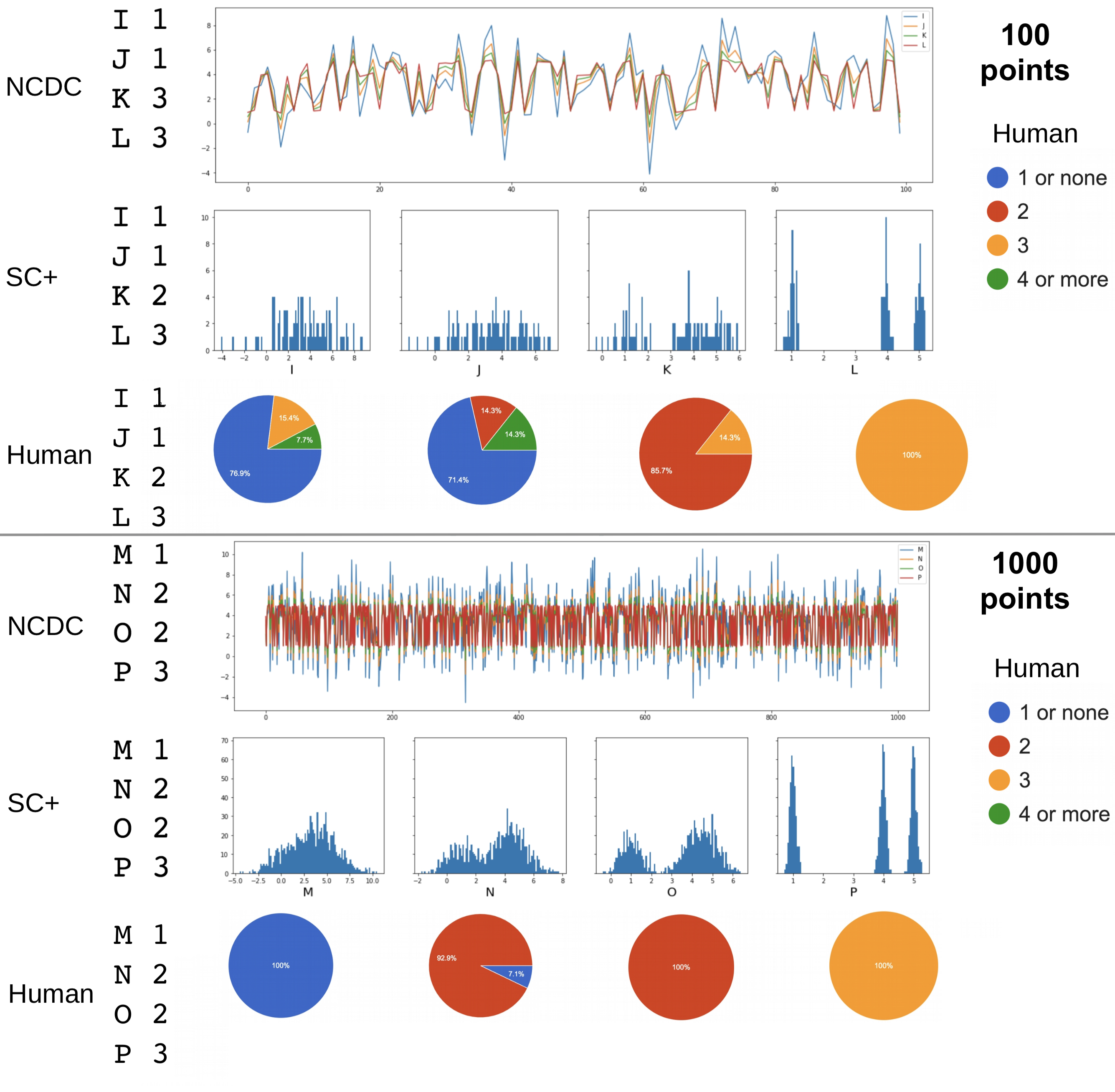}
  \caption{The most revealing results compare human estimates of the number of clusters or distribution modes with the numbers found according to computed metrics such as $NCDC$ and $SSC+$. The upper half is for the case of 100 data points, the lower half is for the case of 1000 data points. The distributions are labeled with the letters I, J, K, L, M, N, O, P. The color graphs represent the data points for all data points corresponding to the respective distributions with the corresponding color legend on the right. The pie charts below the letters correspond to the diversity of human estimates. On the left side, in the three sections for the $NCDC$ metric, the $SC+$ metric, and the most typical human estimate, the selected $K$ numbers are displayed in the corresponding columns next to the letters denoting the distributions.}
  \label{fig5}
\end{figure}

To assess the robustness of our study, we calculated the agreement measures across different ways of determining the number of distribution modes: $NCDC$ versus $SC+$, each person versus each person, $NCDC$ versus people and $SC+$ versus people. The agreement was assessed using two independent metrics, Fleiss' $kappa$ and Krippendorff's $alpha$, according to \cite{Zapf2016}, with the results presented in the Table~\ref{tab1}.

\begin{table}[ht]
\centering
\begin{tabular}{l|r|r|l}
Metric & Fleiss' kappa & Krippendorff's alpha & Kind of agreement\\\hline
$NCDC$ vs. $SC+$    & 0.55   & 0.56  & Moderate agreement  \\
Humans vs. humans   & 0.59   & 0.59  & Moderate agreement  \\
$NCDC$ vs. humans   & 0.47   & 0.48  & Moderate agreement  \\
$SC+$ vs. humans    & \textbf{0.92}  & \textbf{0.92} & Almost perfect agreement
\end{tabular}
\caption{\label{tab1}Assessing the agreement between the numbers of clusters or distribution modes identified by human respondents and based on computable metrics such as $NCDC$ and $SC+$.}
\end{table}

We found that in most agreement estimates, such as $NCDC$ vs. $SC+$, each person vs. each person, $NCDC$ vs. people, all agreement estimates indicate what is called "moderate agreement". Incidentally, we also found that sparser data with fewer data points produce less agreement (see, for example, the structure of the pie charts in Figure~\ref{fig5}).

The most important result observed was that using the $SC+$ metric to determine $K$ has the highest agreement with human estimates, which qualifies as "near perfect agreement". Thus, we conclude that the combined $SC+$ metric, based on the primacy of the Silhouette coefficient and the fallback to "Normalized Centroid Distance times Centroids" in case of uncertainty in the Silhouette coefficient, is an excellent measure for managing human-consistent and human-friendly quantification of numerical data.

\section*{Conclusion}

Based on the presented research, we have reached the following conclusions about how to perform a quantitative assessment of numerical data using one-dimensional clustering to determine whether such assessment is valid at all and, if so, how many data modes or natural clusters of data can be found.

Maximizing the Silhouette coefficient ($SC$) seems more intuitive to humans, but does not allow one to estimate the presence of only one cluster of data, which corresponds to the case where natural clustering is not possible at all due to the presence of only one mode of data in a unimodal distribution such as the normal distribution.

Minimizing the "Normalized Centroid Distance" ($NCD$) - based on the idea of "Minimum Description Length", works for $K=1$, but generally does not match human "reductionist" intuition (reduce the number of entities to a meaningful number) for various distributions, since it tends to create more clusters than are present in the underlying data model.

Minimizing the "Normalized Centroid Distance times Centroids" ($NCDC$) - extends $NCD$ by multiplying it by the number of clusters to penalize creating too many clusters, works for $K=1$ as well as other values of $K$, is more accurate than $NCD$ but less human-intuitive than $SC$ or $SC+$.

Using the combined $SC+$ metric, maximizing $SC$ if it is above the 0.65 threshold (or the Dip Test $pval$ value below 0.5), or minimizing $NCDC$ otherwise, generally seems optimal from a human intuition perspective.

In our future work, we will consider extending and evaluating the presented approach to multi-dimensional data, including applications such as probabilistic causal logic for multi-dimensional financial time series and unsupervised learning for human-interpretable natural language processing models.

\bibliography{sample}

@Article{Boros1997,
author={Boros, Endre
and Hammer, Peter L.
and Ibaraki, Toshihide
and Kogan, Alexander},
title={Logical analysis of numerical data},
journal={Mathematical Programming},
year={1997},
month={Oct},
day={01},
volume={79},
number={1},
pages={163-190},
issn={1436-4646},
doi={10.1007/BF02614316},
url={https://doi.org/10.1007/BF02614316}
}

@Article{Clark1976,
author={Clark, Malcolm W.},
title={Some methods for statistical analysis of multimodal distributions and their application to grain-size data},
journal={Journal of the International Association for Mathematical Geology},
year={1976},
month={Jun},
day={01},
volume={8},
number={3},
pages={267-282},
issn={1573-8868},
doi={10.1007/BF01029273},
url={https://doi.org/10.1007/BF01029273}
}

@INPROCEEDINGS{6137392,
  author={Tewari, Ashutosh and Giering, Michael J. and Raghunathan, Arvind},
  booktitle={2011 IEEE 11th International Conference on Data Mining Workshops}, 
  title={Parametric Characterization of Multimodal Distributions with Non-gaussian Modes}, 
  year={2011},
  volume={},
  number={},
  pages={286-292},
  doi={10.1109/ICDMW.2011.135}}

@book{10.5555/342932,
author = {Kovalerchuk, Boris and Vityaev, Evgenii},
title = {Data mining in finance: advances in relational and hybrid methods},
year = {2000},
isbn = {0792378040},
publisher = {Kluwer Academic Publishers},
address = {USA},
doi = {10.1007/b116453},
}

@article{KOLONIN2022180,
title = {Cognitive Architecture for Decision-Making Based on Brain Principles Programming},
journal = {Procedia Computer Science},
volume = {213},
pages = {180-189},
year = {2022},
note = {2022 Annual International Conference on Brain-Inspired Cognitive Architectures for Artificial Intelligence: The 13th Annual Meeting of the BICA Society},
issn = {1877-0509},
doi = {https://doi.org/10.1016/j.procs.2022.11.054},
url = {https://www.sciencedirect.com/science/article/pii/S1877050922017458},
author = {Anton Kolonin and Andrey Kurpatov and Artem Molchanov and Gennadiy Averyanov}
}

@inproceedings{10.1007/978-3-031-19907-3_4,
author = {Kolonin, Anton and Raheman, Ali and Vishwas, Mukul and Ansari, Ikram and Pinzon, Juan and Ho, Alice},
title = {Causal Analysis of Generic Time Series Data Applied for Market Prediction},
year = {2023},
isbn = {978-3-031-19906-6},
publisher = {Springer-Verlag},
address = {Berlin, Heidelberg},
doi = {10.1007/978-3-031-19907-3_4},
booktitle = {Artificial General Intelligence: 15th International Conference, AGI 2022, Seattle, WA, USA, August 19–22, 2022, Proceedings},
pages = {30–39},
numpages = {10},
location = {Seattle, WA, USA}
}

@article{VITYAEV2015169,
title = {Unified Formalization of "Natural" Classification, "Natural" Concepts, and Consciousness as Integrated Information by Giulio Tononi1},
journal = {Procedia Computer Science},
volume = {71},
pages = {169-177},
year = {2015},
note = {6th Annual International Conference on Biologically Inspired Cognitive Architectures, BICA 2015, 6-8 November Lyon, France},
issn = {1877-0509},
doi = {https://doi.org/10.1016/j.procs.2015.12.191},
url = {https://www.sciencedirect.com/science/article/pii/S1877050915036522},
author = {Evgenii Vityaev}
}

@InProceedings{10.1007/978-3-030-27005-6_11,
author="Glushchenko, Alex
and Suarez, Andres
and Kolonin, Anton
and Goertzel, Ben
and Baskov, Oleg",
editor="Hammer, Patrick
and Agrawal, Pulin
and Goertzel, Ben
and Ikl{\'e}, Matthew",
title="Programmatic Link Grammar Induction for Unsupervised Language Learning",
booktitle="Artificial General Intelligence",
year="2019",
publisher="Springer International Publishing",
address="Cham",
pages="111--120",
isbn="978-3-030-27005-6"
}

@Article{Zapf2016,
author={Zapf, Antonia
and Castell, Stefanie
and Morawietz, Lars
and Karch, Andr{\'e}},
title={Measuring inter-rater reliability for nominal data -- which coefficients and confidence intervals are appropriate?},
journal={BMC Medical Research Methodology},
year={2016},
month={Aug},
day={05},
volume={16},
number={1},
pages={93},
issn={1471-2288},
doi={10.1186/s12874-016-0200-9},
url={https://doi.org/10.1186/s12874-016-0200-9}
}

@Inbook{Jin2010,
author="Jin, Xin and Han, Jiawei",
title="K-Means Clustering",
bookTitle="Encyclopedia of Machine Learning",
year="2010",
publisher="Springer US",
address="Boston, MA",
pages="563--564",
isbn="978-0-387-30164-8",
doi="10.1007/978-0-387-30164-8\_425",
}

@INPROCEEDINGS{9356727,
  author={Deng, Dingsheng},
  booktitle={2020 7th International Forum on Electrical Engineering and Automation (IFEEA)}, 
  title={DBSCAN Clustering Algorithm Based on Density}, 
  year={2020},
  volume={},
  number={},
  pages={949-953},
  doi={10.1109/IFEEA51475.2020.00199}}

@article{10.1214/aos/1176346577,
author = {J. A. Hartigan and P. M. Hartigan},
title = {{The Dip Test of Unimodality}},
volume = {13},
journal = {The Annals of Statistics},
number = {1},
publisher = {Institute of Mathematical Statistics},
pages = {70 -- 84},
year = {1985},
doi = {10.1214/aos/1176346577},
URL = {https://doi.org/10.1214/aos/1176346577}
}

@inbook{Bauer_2023,
author={Bauer, Lena G. M. and Leiber, Collin and Böhm, Christian and Plant, Claudia},
title = {Extension of the Dip-test Repertoire - Efficient and Differentiable p-value Calculation for Clustering},
booktitle = {Proceedings of the 2023 SIAM International Conference on Data Mining (SDM)},
chapter = {},
pages = {109-117},
doi = {10.1137/1.9781611977653.ch13},
publisher={Society for Industrial and Applied Mathematics},
year={2023},
eprint = {https://epubs.siam.org/doi/pdf/10.1137/1.9781611977653.ch13}
}

@INPROCEEDINGS{4426662,
  author={Aranganayagi, S. and Thangavel, K.},
  booktitle={International Conference on Computational Intelligence and Multimedia Applications (ICCIMA 2007)}, 
  title={Clustering Categorical Data Using Silhouette Coefficient as a Relocating Measure}, 
  year={2007},
  volume={2},
  number={},
  pages={13-17},
  doi={10.1109/ICCIMA.2007.328}}

@InProceedings{10.1007/978-3-642-23851-2_9,
author="Georgieva, Olga
and Tschumitschew, Katharina
and Klawonn, Frank",
editor="K{\"o}nig, Andreas
and Dengel, Andreas
and Hinkelmann, Knut
and Kise, Koichi
and Howlett, Robert J.
and Jain, Lakhmi C.",
title="Cluster Validity Measures Based on the Minimum Description Length Principle",
booktitle="Knowledge-Based and Intelligent Information and Engineering Systems",
year="2011",
publisher="Springer Berlin Heidelberg",
address="Berlin, Heidelberg",
pages="82--89",
isbn="978-3-642-23851-2"
}

@misc{pagnoni2024bytelatenttransformerpatches,
      title={Byte Latent Transformer: Patches Scale Better Than Tokens}, 
      author={Artidoro Pagnoni and Ram Pasunuru and Pedro Rodriguez and John Nguyen and Benjamin Muller and Margaret Li and Chunting Zhou and Lili Yu and Jason Weston and Luke Zettlemoyer and Gargi Ghosh and Mike Lewis and Ari Holtzman and Srinivasan Iyer},
      year={2024},
      eprint={2412.09871},
      archivePrefix={arXiv},
      primaryClass={cs.CL},
      url={https://arxiv.org/abs/2412.09871}, 
}

@InProceedings{10.1007/978-3-031-44865-2_1,
author="Kolonin, Anton",
editor="Kryzhanovsky, Boris
and Dunin-Barkowski, Witali
and Redko, Vladimir
and Tiumentsev, Yury
and Klimov, Valentin",
title="Evolution of Efficient Symbolic Communication Codes",
booktitle="Advances in Neural Computation, Machine Learning, and Cognitive Research VII",
year="2023",
publisher="Springer Nature Switzerland",
address="Cham",
pages="3--12",
isbn="978-3-031-44865-2"
}

@Article{Sierra-Sosa2023,
author={Sierra-Sosa, Daniel
and Pal, Soham
and Telahun, Michael},
title={Data rotation and its influence on quantum encoding},
journal={Quantum Information Processing},
year={2023},
month={Jan},
day={23},
volume={22},
number={1},
pages={89},
issn={1573-1332},
doi={10.1007/s11128-023-03837-1},
url={https://doi.org/10.1007/s11128-023-03837-1}
}

\end{document}